\documentclass[12pt]{article}
\usepackage[top=1in, bottom=1in, left=1.in, right=1.in]{geometry}
\usepackage[english]{babel}
\usepackage{atbegshi,cite}
\usepackage{amsmath,amssymb,amsbsy,amstext, amsthm, simplewick}
\usepackage{hyperref}
\usepackage{graphicx}
\usepackage{amsfonts}
\usepackage[small]{caption}
\usepackage{upgreek}
\usepackage[titletoc]{appendix}
\usepackage{setspace}
\usepackage[dvipsnames]{xcolor}
\usepackage{slashed}
\usepackage{soul}

\newcommand{\newc}{\newcommand}
\newc{\be}{\begin{equation}}
\newc{\ee}{\end{equation}}
\newc{\fpi}{f_{\pi}}
\newc{\etap}{\eta^{\prime}}
\newc{\llll}{\langle\lambda\lambda\rangle}
\newc{\FFd}{F^a\tilde F^a}
\newc{\qbar}{{\overline q}}
\newc{\TR}{{\rm Tr}}
\newc{\Kahler}{K\"ahler }
\newc{\Zbb}{{\mathbb Z}}
\newc{\Rt}{{\mathbb R}^3}
\newc{\Rf}{{\mathbb R}^4}
\newc{\So}{{\mathbb S}^1}
\newc{\zt}{{\mathbb Z}_2}
\newc{\RtSo}{{\mathbb R}^3\times{\mathbb S}^1}
\newc{\scriminus}{{\cal I}^-}
\newc{\scriplus}{{\cal I}^+}
\newc{\mpl}{M_p}
\newc{\Ricci}{\mathcal{R}}
\newc{\bv}{\phi}
\newc{\calU}{{\cal U}}
\newc{\calK}{K}
\newc{\calUi}{{\cal U}^{-1}}
\newc{\calG}{{\cal G}}
\newc{\calM}{{\cal M}}
\newc{\calL}{{\cal L}}
\newc{\calO}{{\cal O}}
\newc{\calR}{{\cal R}}
\newc{\calQ}{{\cal Q}}
\newc{\calI}{{\cal I}}
\newc{\calOb}{{\cal O}^\dagger}
\newc{\hphi}{{\hat\phi}}

\newc{\tb}[1]{{\bf\color{MidnightBlue}TB:~#1}}
\newc{\pd}[1]{{\color{blue}#1}}
\newc{\pdq}[1]{{\color{red}#1}}

\theoremstyle{plain}
\theoremstyle{plain} 
\theoremstyle{plain} 
\theoremstyle{plain}
\theoremstyle{plain}
\theoremstyle{plain}

\renewcommand{\title}[1]{{\Large\bf\flushleft{#1}}\vspace*{3ex}\\}
\renewcommand{\author}[2]{{\noindent\hspace*{2.5em}\large#1}
                     \footnote{Electronic mail: $\mathtt{#2}$}\\}

\begin{document}
\begin{titlepage}
\begin{flushright}
{\large 
~\\
}
\end{flushright}

\vskip 2.2cm

\begin{center}

{\large \bf JT Gravity Coupled to Fermions}

\vskip 1.4cm

{{Tom Banks}$^{(a)}$,  {Patrick Draper}$^{(b)}$, and  {Bingnan Zhang}$^{(a)}$}
\\
\vskip 1cm
{$^{(a)}$ Department of Physics and NHETC, Rutgers University, Piscataway, NJ 08854}\\
{$^{(b)}$ Department of Physics, University of Illinois, Urbana, IL 61801}
\vspace{0.3cm}
\vskip 4pt

\vskip 1.5cm

\begin{abstract}
We argue that two-dimensional dilaton gravity models can all be derived from an analog of Jacobson's
covariant version of the first law of thermodynamics.  We then specialize to the JT gravity model and couple it to  massless fermions. This model is exactly soluble in quantum field theory, and we present a new derivation of that result.

 The field theory model violates two principles one might want to impose on a quantum theory of gravity describing the near horizon region of an extremal charged black hole in four dimensions: finiteness of the entropy for finite causal diamonds, and the absence of global conservation laws.  It preserves an infinite number of conservation laws that one would have expected to be violated, since the fermion state on each side of the $AdS_2$ wormhole is unavoidably thermal.  We describe a cutoff version of the model, with extra interactions, which cures these difficulties.   Our UV completion of the model depends on the AKK\cite{AKK} map of non-relativistic fermions in an inverted oscillator potential to Weyl fermions in Minkowski space.  We argue that gauging the $Z_2$ symmetry of the oscillator model, using a density matrix with temperature that depends on the oscillator coordinates, and inserting chaotic interactions at (almost) infinite  oscillator coordinate, we obtain a model with  properties expected of quantum gravity in the near horizon region of an extremal charged black hole in four dimensions.\end{abstract}

\end{center}

\vskip 1.0 cm

\end{titlepage}
\setcounter{footnote}{0} 
\setcounter{page}{1}
\setcounter{section}{0} \setcounter{subsection}{0}
\setcounter{subsubsection}{0}
\setcounter{figure}{0}



\section{Introduction}
Models of dilaton gravity in two dimensions have given valuable insights into the properties of quantum theories of gravity. In recent years, the Jackiw-Teitelboim (JT) model~\cite{JT} has received renewed attention because it appears to be dual to a class of non-gravitational models of interacting fermions~\cite{SYK}.  One of the problems with this duality is that, using the conventional logic of AdS/CFT, the SYK model does not appear to be dual to an approximately local effective field theory in the bulk.  The purpose of the present paper is to construct a model dual to 
JT gravity coupled to  free massless fermions, which exhibits local physics explicitly.

JT gravity can be related to 4D Einstein-Maxwell theory by dimensional reduction on a sphere and linearization of the dilaton. Solutions of the theory can be thought of as describing the physics of the near-horizon limit of 4D extremal Reissner-Nordstrom (RN) black holes. The classical black hole solutions possess a nonzero entropy at zero temperature without supersymmetry, and the SYK model purports to resolve that vacuum degeneracy.  The $AdS_2$ throat of the black holes has macroscopic radius of curvature for large charge, and one would have expected the dual model to be well-approximated by a local 2D effective field theory.  The model we construct has that property. It is similar in spirit to the UV-completion of the CGHS model proposed by one of the authors~\cite{cghsrst}\cite{lindil1}\cite{lindil2}, 
although we will see the physics is quite different. The action we consider is:
\begin{align}
{\cal S}=\int d^2x\,\sqrt{-g} \left(N S(R + 1) + i\bar{\psi}_J e^m_a \gamma^a (i\partial_m - \omega_m)\psi_J \right),
\label{eq:lag}
 \end{align} 
 where  the metric, zweibein, and spin connection $\omega$ are related in the usual fashion.  $N$ is the number of independent fermion fields. We have chosen units of length such that the radius of curvature is equal to $1$. The dilaton $S$ will have an entropic interpretation, and we will refer to it as the entropy field below. We will mostly set $N = 1$.  The model is exactly soluble for any $N$, and one finds that higher order corrections to the $1/N$ expansion vanish. Previous studies on the same topic of JT gravity coupled to conformal field theories include~\cite{trivedi}.

The model~(\ref{eq:lag}) is a renormalizable quantum field theory, though not all of the couplings that one might have expected from dimensional analysis are present.  In particular, there are no couplings between the $S$ field and the fermions.  In~\cite{cghsrst}, this was a choice motivated by viewing the bosonized fermions as the remnants of Ramond-Ramond fields in a string compactification.  In the model studied here, it is an exact property that no such couplings are induced by quantum loops.  We may view this as a consequence of the fact that the metric-dilaton system has no independent field degrees of freedom.  Thus, although we formulate the problem as an interacting field theory, it is really a free field theory coupled to a single bosonic quantum degree of freedom, which is topological in the sense that it describes global properties of the space-time/dilaton configuration.
We will view the global solution space of JT gravity as establishing a set of parameters that define a model, rather than a set of variables that must be quantized as in\cite{hj}. The real reason this model is simpler than that of CGHS is that the metric is a frozen classical field and the dynamical fermions couple only to the metric, even after renormalization effects are taken into account.  We will discuss this in more detail below.

We have referred above to the interpretation of $S$ as an entropy. We now describe the sense in which all dilaton gravity models in two dimensions are related to hydrodynamics, using Jacobson's relation between gravity and thermodynamics~\cite{ted}. In somewhat anachronistic language, Jacobson's result in higher dimensions is that the null-null components of Einstein's equations are nothing but the first law of thermodynamics $dE = T dS$, applied locally to the area law for the entropy of a causal diamond in Lorentzian spacetime.  In this derivation, energy is interpreted as the energy measured along a maximally accelerated Unruh trajectory, with inflection point on the holographic screen\footnote{The maximal $d - 2$ volume leaf of a null foliation of the diamond boundary.} of the diamond.  Maximal acceleration means that the temperature goes to infinity as the trajectory becomes null, which tells us that we are discussing the maximum possible entropy in the diamond. In quantum mechanics the maximum entropy is the logarithm of the dimension of the Hilbert space associated with measurements done in the diamond.  (In the quantum theory, the infinite acceleration is cut off by the Planck scale, and the system has a density matrix that exhibits entropy fluctuations.)

Jacobson's derivation fails in two dimensions, because diamond boundaries are points and have no ``transverse area."   Instead, we simply postulate the existence of a scalar entropy function $S$. The values of $S$ at the spatial tips of diamonds determine the entropy of the Hilbert space describing the diamond in a two dimensional model of quantum gravity.   The heat flux through a horizon $H$ is~\cite{ted}
\begin{align}
    dE \propto \int_H d\lambda \,\lambda  T_{ab}k^a k^b
\end{align}
where $\lambda$ is an affine parameter and $k^a$ is tangent to the horizon. We write the entropy flux as
\begin{align}
    dS \propto \int_H d\lambda \,  (\nabla_a  S)k^a ~ \propto~ -\int_H d\lambda \,\lambda  (\nabla_a \nabla_b S)k^a k^b
\end{align}
(where $S$ on the right-hand side is the dilaton) and impose the first law $dE=TdS$. Then, since the space-like boundaries of a diamond can be any pair of points in two dimensions, we obtain two equations relating $S$ to null energy. For a suitable dilaton normalization they take the form:
\begin{align}
&\nabla_+^2 S = T_{++} \nonumber\\
&\nabla_-^2 S = T_{--} .
\label{dilatonconstraints}
\end{align}
Note that if we work with coordinates that have dimensions of length these equations are dimensionally correct for an energy density and an entropy.  The interpretation of entropy as a scalar function can be motivated by dimensional reduction, and has also been validated in linear dilaton models\cite{cghsrst}\cite{lindil1}\cite{lindil2}, where the variation in area comes from variation of the Planck scale at fixed string scale.

The stress energy tensor is, by definition, a conserved symmetric tensor, so the covariant version of these equations must be
\begin{equation} \nabla_m \nabla_n S + g_{mn} U(S) = T_{mn}  \label{eq:Seq}\end{equation}
where $U$ is some function of $S$ and its derivatives.
Covariant conservation of the stress tensor is equivalent to the requirement that this tensor equation arises from the metric variation of a generally covariant action.  This action has a derivative expansion if we assume fields are slowly varying, so it can be written
\begin{equation} {\cal S} = \int \sqrt{-g} [S R + F(S) g^{mn} \nabla_m S \nabla_n S - V(S) + g_{mn}T^{mn}] . \end{equation}  (The metric variation of the $SR$ term gives rise to the $\nabla_m \nabla_n S$ term in Eq.~(\ref{eq:Seq}) and the variation of the $S$ kinetic term contributes to the $U(S)$ term.) It is important to emphasize that in this phenomenological hydrodynamic derivation of the equations, $T_{mn}$ is the expectation value, in some state, of the full stress tensor of the underlying model.  We can do a field redefinition of the metric, with the form of an $S$-dependent Weyl transformation, which reduces one or the other of the functions $F$, $V$ to some special form, like a constant.  So different versions of the Lagrangian depend on a single function of $S$ and a few constants. Note that using Jacobson's Principle as the origin of these equations,  $S$ does not couple to the stress tensor. In both the CGHS model, and previous studies of JT gravity coupled to conformal matter, a similar restriction was motivated by  different arguments.

The extra freedom, compared to higher dimensional models, is a consequence of the fact that there is a unique pair of null directions for any causal diamond.  Another related feature is that, because the holographic screens are points, there is no notion of localization of entropy on parts of the screen.  As we will see, this is reflected in the behavior of quasi-normal modes, the high energy behavior of the entropy, and the details of scrambling in the underlying quantum mechanics.  It will also tell us that the JT model that we study is not a good model of the classical behavior of $4$-dimensional near-extremal Reissner-Nordstrom black holes.  It represents only the dynamics of the nearly degenerate ground states implied by the zero temperature entropy of these models.

The dilaton expectation value in any state satisfies Eq.~(\ref{dilatonconstraints}), showing that variations of it behave like variations of a local entropy density.   Our quantization procedure will resolve another puzzle about the view of $S$ as a quantum field.   The energy in the first law of thermodynamics is the expectation value of the {\it full} Hamiltonian of the quantum theory.   Adding extra pieces to the Hamiltonian involving the $S$ field and the metric does not sound consistent.   This observation suggests that it  is incorrect to treat
JT gravity as a stand-alone quantum system
. Instead we'll find that the metric is a completely classical field and the fluctuating part of S is a constrained field whose quantum algebra is entirely determined by that of the fermion stress tensor.

This work is organized as follows.

In Sec.~\ref{sec:classical} we study the field equations. The geometry is fixed to be locally AdS, and we emphasize that the static coordinate system has a preferred status, giving a time-independent  solution for the dilaton. We discuss the role of $SL(2,R)$ symmetry breaking by the $S$ background. 

In Sec.~\ref{sec:quantum} we quantize the theory. Following the quantization procedure due to Schwinger, we promote fields to operators such that the equations of motion follow from the Heisenberg equations, and the constraint equations are generated as consistent operator constraints. We find that the dilaton equation follows  from  the fermion equations of motion and the constraints, while the metric remains a classical background and is not quantized. Thus, quantizing the fermions as free fields leads to a complete quantization of the full set of equations of motion and constraints. 

The Schwinger method can be carried out formally in any coordinate system for the classical gravitational field.  However, it has a time independent Hamiltonian implementation only in static coordinates.  These coordinates cover only half the global time interval of the full AdS space.    Using static coordinates we can make a Euclidean continuation.  As usual, the presence of Killing horizons makes the Euclidean manifold singular unless Euclidean time is periodic.  There are multiple possibilities for the topology of manifolds that heal the singularities.  Following Coleman\cite{coleman} and Saad {\it et. al.}\cite{sss} we argue that all but the simplest disconnected topology correspond to averaging the model over random couplings.  Restricting to the disconnected topology we obtain a thermofield double interpretation for the Euclidean path integral.  Euclidean methods confirm that the metric is a frozen classical variable, and also enable us to study the renormalization properties of our model.  We find infinite renormalizations of the two dimensional cosmological constant and an infinite additive shift of the $S$ field.  Euclidean methods also reproduce the conformal anomaly, which is derived by point splitting regularization of the stress tensor in Schwinger's method.

In Section~\ref{sec:qnm} we study semi-classical solutions of the equations for the bosonic potentials of fermion currents, which we interpret as quasi-normal modes of the $AdS_2$ horizon.  Their most important property is that a concentration of energy density, initially localized a finite distance from the horizon and traveling towards the boundary, returns to an infinitesimal neighborhood of the horizon in a finite static time, and remains there.  Hawking radiation implies that there will always be some probability to find fermions far from the horizon, but in equilibrium states it will always be small.  We also find that the value of $S$ near the horizon can be changed by quasi-normal mode solutions, so that the model can make transitions between different static solutions. 

In Section~\ref{sec:IHO} we introduce an exact quantum version of this semi-classical model.  The aim is to provide a model that exhibits local physics in the throat of an extremal 4d black hole.  That throat has a radius of curvature much larger than all microscopic length scales.  As mentioned above, the SYK models do not have an obvious description of local physics, and a naive interpretation of them using the AdS/CFT dictionary indicates that the AdS radius is ``of order the string scale."  There is no large gap in the spectrum of conformal dimensions. Instead, we proceed as follows.
In the field theory~(\ref{eq:lag}), the fermions interact only with the metric, so they are free  on the wormhole geometry.  The geometry is conformal to flat space, so apart from the conformal anomaly and boundary conditions, the problem reduces to free massless fermions in Minkowski space. We then use the AKK transformation (a local map between the spatial Minkowski coordinate to the ``$u$" coordinate of an inverted oscillator with $H = [uv + vu]/2$) to map the free relativistic fermions to non-relativistic (NR) fermions in an inverted oscillator.    All calculations in the relativistic model can be written in terms of the non-relativistic problem.   This mapping was used to make a quantum model of linear dilaton gravity in\cite{lindil1}\cite{lindil2}, but we have to use different boundary conditions on the oscillator model to reproduce the expected properties of quantum JT gravity.

Using the WKB approximation in the NR inverted oscillator problem, we argue that the states of interest correspond to certain density matrices of the NR fermions by comparing to the properties of the states of the relativistic fermions in the wormhole geometry.   In particular, the relation between entropy in a causal diamond with one spacelike boundary anchored at infinity, calculated from the classical dilaton profile, and the static transit time in that diamond, can be mapped to the NR fermion model  (this  tells one how to map the oscillator coordinate onto the spatial coordinate in the static conformal coordinate system).  Also, the two calculations agree that the equilibrium states of the model have all fermions clustered near the horizon (which maps to infinite oscillator coordinate).  In the geometric theory it is inappropriate to think of the observables as living on the infinite boundary. 

In order to get finite entropy and destroy the integrability of the model, we cut off the infinite oscillator coordinate and impose Dirichlet boundary conditions there.  We also impose a UV cutoff on the single particle energies.  The UV cutoff has a negligible effect on the density matrices since very high energy states are exponentially suppressed.  Finally, we place SYK interactions at the oscillator coordinate cutoff.  Therefore, in our density matrices, the model essentially reduces to the SYK model, but it also has non-equilibrium states which describe relativistic fermions propagating in $AdS_2$ far from the horizon.  We take a single instantiation of SYK couplings, so the model is pure quantum mechanics and its Euclidean path integral treatment does not involve wormholes.  More details of the map and boundary conditions on the oscillator, appropriate to the $AdS_2$ geometry, will be given below.  As in the linear dilaton case, the connection of the classical geometry to the actual quantum model is evident in the WKB approximation to the non-relativistic problem, and is distorted by quantum corrections.

\section{Classical Solutions}
\label{sec:classical}

The classical equations of motion of JT gravity coupled to free massless Dirac fermions (cf. Eq.~(\ref{eq:lag})) are
\begin{align}
 (\nabla_m \nabla_n &S - \frac{1}{2}g_{mn} S) = \frac{1}{N}T_{mn}\nonumber\\
& R + 1 = 0 \nonumber\\
 &e^m_a \gamma_a D_m \psi = 0 . 
 \label{eq:EOM}
\end{align}
$T_{mn}$ is the stress tensor of the  fermions. In the large $N$ limit an $\calO(1)$ value of $S$  is achieved when the stress tensor expectation value is order $N$. An $\calO(1)$ value of $S$ corresponds to a large entropy,  which is the case for states studied by conventional large $N$ methods.   Since the model is exactly soluble, we will not really use large $N$ techniques and will, for the most part, set $N = 1$.   We will however assume that the constant in the solution of the dilaton equation of motion is large.

\subsection{Geometry}

The second equation in~(\ref{eq:EOM}) reflects the fact that $S$ acts as a Lagrange multiplier, fixing the metric to be locally AdS$_2$.
 There are four interesting coordinate systems on $AdS_2$ relevant for our purposes: global coordinates, static coordinates, and conformally flat versions of each.   
 
 The conventional global coordinates are 
\begin{equation} ds^2 = - 2 (1 + x^2) d\tau^2 + \frac{2dx^2}{1 + x^2} . \end{equation}  
These dimensionless coordinates cover the whole manifold and its entire extended Penrose diagram (the universal cover of the hyperboloid in a flat Minkowski 3-space with two time directions), if we allow $\tau$ to take on all real values.   

The global conformal coordinates are $(\tau,y)$, where 
$x = \tan(y)$, so that $y$ ranges from $-\pi/2$ to $\pi/2$. In these coordinates conformal flatness is manifest,
\begin{align}
g_{mn} &= e^{2\sigma(y)} \eta_{mn}\nonumber\\
e^{2\sigma(y)}&=2\sec^2(y)=2(1+x^2).
\end{align}

The static coordinates $(t,r)$ are related to the global coordinates by
\begin{align}
&x=\pm \sqrt{(r/r_s)^2-1}\cosh(r_s t)\nonumber\\
&\tau = \arccos\left(\frac{1}{\sqrt{1+(1-(r_s/r)^2)\sinh^2(r_s t)}}\right).
\end{align}
The metric takes the form 
\begin{align}
ds^2=-2(r^2-r_s^2)dt^2+\frac{2dr^2}{r^2-r_s^2}.
\end{align}
The static coordinates are restricted to $r^2\geq r_s^2$, and they cover two causal diamonds with horizons at $r=\pm r_s$. In global conformal coordinates, the point $y=0$ maps to {\emph{both}} horizons. The mapping is singular at these points.

We define a  static conformal coordinate $s$ by
\begin{align}
r=r_s\coth(r_s s).
\end{align}
In static conformal coordinates the metric is  
\begin{align}
g_{mn} &= e^{2\sigma(s)} \eta_{mn}\nonumber\\
 e^{2\sigma(s)} &= 2 r_s^2 \,{\rm csch}^2(r_s s) = 2(r^2-r_s^2).
\end{align}

It is also convenient to introduce light front  coordinates in both the global-conformal and  static-conformal cases,
$y^\pm = \tau\mp y$,  $s^\pm = t\mp s$.
 
 \subsection{$S$  Solutions}

 The trace of the $S$ equation in Eq.~(\ref{eq:EOM}) is the massive Klein-Gordon equation with source,
 \begin{align}
 \nabla^2 S -  S =g^{mn}T_{mn}.
 \end{align}
This is an exact operator equation. The trace of the stress tensor is given exactly by the conformal anomaly
   \begin{equation} g^{mn}T_{mn} = -\frac{1}{24}  R. \end{equation}  
   We can write the solution for $S$ as
   \begin{equation}  S = S_h +   \int G(t,x , t^{\prime} x^{\prime}) dt^{\prime} dx^{\prime} , \label{eq:Sgeneralsol}\end{equation} 
   where $S_h$ a homogeneous solution, and $G$ is a massive Klein-Gordon Green's function discussed further below.
   
If the gravitational bulk field theory is dual to a quantum theory with a time-independent Hamiltonian, then we need coordinates in which the $S$ field has a time-independent solution, at least for some choice of the quantum state. This turns out to be restrictive.

    The second term in Eq.~(\ref{eq:Sgeneralsol}) will be $t$-independent, for some time coordinate $t$, if $\partial_t$ is any Killing vector in AdS$_2$. There is only one static solution of the homogeneous equation that is well behaved on the boundary, and it is static in static coordinates:
    \begin{equation} S_h = -\mu^2 r. \label{Shsol}\end{equation} 
    
In contrast, in global coordinates, $S_h \propto\sqrt{1+x^2} \cos (\tau)$. In other words, the solution is not static with respect to $\tau$, and the global timelike Killing vector $\partial_\tau$ cannot be identified with the time of a time independent quantum system.

We have chosen the overall sign in Eq.~(\ref{Shsol})   so that the  homogeneous solution for $S$ is negative. This choice is based on our interpretation of $S$ as the entropy of a certain one parameter family of causal diamonds, which is described further below. We will also comment further in Sec.~\ref{sec:RNextremal} about the relationship of this choice to the 4D extremal RN solution. $S$ also has a constant term undetermined by the field equations. The constant  is shifted by an infinite positive renormalization in quantum field theory.  We will fix it to determine the total entropy of the system in the UV completion of the model.
   
Note also from the global solution that $S_h\rightarrow0$ at $\tau = \pm\pi/2$, before the complete (multiply-connected) AdS$_2$ manifold is covered.  This ``zero-entropy point" should be considered a space-like singularity, like the zero area singularities of higher-dimensional black holes.  In static coordinates, this singularity appears at $r = 0$, near the Killing horizons of the two asymptotic regions $r>r_s$.  From the point of view of classical field theory, $S = 0$ is a singularity because the classical kinetic term vanishes at that point.  In our UV completion of the model, the boundary will be a finite space-like distance from the horizon and there will be a zero of the entropy function at the finite boundary from which all of our causal diamonds are measured.  Reflecting boundary conditions on the finite boundary are perfectly consistent.

We see that, once the scalar field configuration is taken into account, only the wormhole geometry is consistent with the interpretation of $S$ as entropy and/or the absence of classical singularities in space-time.   We note that the interpretation of $S = 0$ as a singularity is standard procedure in CGHS and related models. 

The $AdS_2$ geometry has an $SL(2,R)$ symmetry that is useful for understanding some of the physics, see for example~\cite{kitaevsuh}. However, 
the $S$ solution breaks the $SL(2,R)$ invariance, preserving only the time translation symmetry in static coordinates, which is one of the non-compact generators of $SL(2,R)$.  The compact generator is $\tau$ translation and the other non-compact generator is (asymptotically) the simultaneous rescaling of $r$ and $t$ in opposite directions.  Under this transformation the parameters of the $S$ solution are re-scaled.  $r_s$ scales like $r$ and $S_0$ is invariant.  Thus, only one non-compact generator of $SL(2,R)$ is preserved by the classical solution of JT gravity.  One should not expect the quantum theory dual to this geometrical picture to be $SL(2,R)$ invariant.

\section{Quantization}
\label{sec:quantum}
We will follow an approach to quantization due to Schwinger\cite{Schwinger}.  The basic method is to start from a system of classical field equations, promote various fields to operators, with commutation relations determined by requiring (1) consistency with the general principles of quantum mechanics (unitarity), and (2) that the Hamiltonian operator generating time translations generates both the equations of motion as the Heisenberg equations and the constraint equations as consistent operator constraints.  After that we will turn to Euclidean functional integral quantization and show that if we restrict the topology of the Euclidean metric, we obtain results consistent with Schwinger's method.  

\subsection{Metric and Dilaton}
As discussed above, the metric equation has a unique local solution, and  the  wormhole geometry is the unique global solution for which the static time is well-defined.

The other equations of motion are, as we will see, compatible with taking the metric to be a classical background. In other words, we will not promote the metric to an operator in the quantization.

It is convenient to rewrite the equations~(\ref{eq:EOM})  in conformal gauge with light front coordinates, where the metric is 
\begin{align}
g_{+-} &= - \frac{1}{2}e^{2\sigma} \nonumber\\
g_{++} &= g_{--} = 0.
\end{align}
The equations of motion are
\begin{align}
& \partial_+ \partial_- \sigma + \frac{1}{8} e^{2\sigma} = 0 \nonumber\\
& \partial_+ \partial_- S + \frac{1}{4} e^{2\sigma} S  =T_{+-} \nonumber\\
&\partial_+ \psi_+ = \partial_- \psi_- = 0. 
\end{align}
The remaining two $S$ equations are constraint equations, 
\begin{align}
 \nabla_+^2 S - T_{++} &= 0\nonumber\\
 \nabla_-^2 S - T_{--} &= 0  .
 \end{align}
These linear equations can be solved as operator equations without ordering ambiguities, since we have hypothesized that the metric is a classical function.    The equations of motion are consistent with the constraints because of the covariant conservation law satisfied by $T_{nm}$, which determines the right-moving part of the left-moving stress tensor, and vice versa, in terms of $T_{+-}$.    Classically $T_{+-} = 0$, but this is inconsistent with the fermion quantization rules above.   Instead we have the conformal anomaly
\begin{equation} g^{+-}T_{+-} = \frac{1}{48}  R = -\frac{1}{48}  .  \end{equation} 
This is a c-number source for the $S$ equations.  Since the curvature is constant, the trace piece of the stress tensor is covariantly conserved by itself.

Quantization of these equations means finding a set of operators and a Hamiltonian generating time translation in static time, such that (1)  the constraints can be solved on the Hilbert space of the system, and (2) the equations of motion are the Heisenberg equations of motion, at least on the constrained subspace. The field $\sigma$ is thus a classical field, and in global conformal gauge the solution is $e^{2\sigma}=2\sec^2{y}$. We still have the residual gauge freedom of adding an arbitrary solution of the massless Klein-Gordon equation to $\sigma$.  The constraints are the statement that states are invariant under this transformation.

We split $S$ into a c-number part and an operator part, and denote the operator part as $\hat S$. Thus $\hat S$ obeys the constraint equations 
\begin{align}
& \nabla_+^2 \hat{S} = \frac{1}{N}{T}_{++} \nonumber\\
& \nabla_-^2 \hat{S} = \frac{1}{N}{T}_{--} 
\end{align}
and its classical equation of motion is
\begin{align}
\nabla^2 \hat{S} - \hat S = 0  \label{classicalSeom}
\end{align}
where the Laplacian acting on a scalar is $\nabla^2\hat S = -4 e^{-2\sigma}\partial_+\partial_- \hat S$. These linear equations determine $\hat S$ inside any causal diamond in terms of the values of the stress tensor operators on the past boundaries of that diamond.\footnote{We give the full set of semiclassical equations obeyed by the $S$ field, including quantum corrections to the stress tensor, when we discuss shock waves below.}

It remains to show that the massive KG equation for $\hat S$, Eq.~(\ref{classicalSeom}), follows from the fermion equations of motion, which are the Heisenberg equations following from the fermion Hamiltonian.  To do so  we use conservation of the fermion stress tensor, which holds as an operator equation,
\begin{equation} \nabla_+ T_{--} = \nabla_-  T_{++} = 0 . \end{equation} Therefore, the constraints imply $\nabla_- \nabla_+^2 \hat S= \nabla_+ \nabla_-^2\hat S =0$, or
\begin{align}
 [\nabla_+, \nabla_-] \nabla_{\pm} \hat S &= \pm \left(\nabla_+ \nabla_- + \nabla_- \nabla_+\right) \nabla_{\pm} \hat S \nonumber\\
 &= \pm g_{+-} \nabla^2 \nabla_\pm \hat S.
 \end{align}
 But the commutator is 
 \begin{align}
[\nabla_+, \nabla_-] \nabla_{\pm} \hat S &=  R^a{}_{\pm -+} \nabla_{a} \hat S \nonumber\\
 &= \pm \frac{1}{2} g_{+-}\nabla_\pm \hat S .
  \end{align}
  So
  \begin{align}
  \nabla^2 \nabla_\pm \hat S = \frac{1}{2} \nabla_\pm \hat S.
  \end{align}
  Furthermore, it is straightforward to check that $[\nabla^2,\nabla_\pm]\hat S = -\frac{1}{2} \nabla_\pm \hat S$, so
  \begin{align}
  \nabla_\pm \nabla^2\hat S = \nabla_\pm \hat S.
  \end{align}

This shows that $S$ itself satisfies the massive KG equation up to a possible constant shift. 
We can take the shift to be a c-number, without violating any of the equations.  We will see below that this shift is required in order to absorb an infinite additive renormalization of $S$ induced by the fermions. 

We conclude that simply quantizing the fermions as free fields on a half infinite interval in static conformal coordinates leads to a complete quantization of the full set of equations of motion and constraints, consistent with unitarity and positivity. 
The only ambiguity has to do with boundary conditions at infinity, and we will return to discuss this below.  There is no need for Fadeev-Popov ghosts, since the metric is not a quantum variable.\footnote{Note that the same remark holds true in the Euclidean path integral discussion of\cite{sss}\cite{sw}. The integral over metrics is reduced to an integral over the moduli space of Riemann surfaces.  The boundary Schwarzian field can be viewed as a remnant of the dilaton, $S$.}  The field $S$ has both a quantum and a classical piece, but the quantum piece is entirely determined by the constraints in terms of fermion operators.  Note that while we have chosen a conformal gauge for convenience, the above derivations could be carried out in any coordinate system.  The Hilbert space is completely defined by the quantization of the fermion field, and we can choose Hartle-Hawking boundary conditions at infinity in static conformal coordinates.

\subsection{Euclidean Path Integral Quantization}
We follow Ref.~\cite{hj,jw} and define the Euclidean path integral by making a Euclidean continuation of the static coordinate time.  We will also follow\cite{sss} and assume that the Euclidean $S$ field is a non-negative classical background plus an imaginary piece which is integrated over.  As a consequence, the Euclidean metric must have constant negative curvature, which means that its only fluctuations are in its topology and the moduli space of metrics of fixed topology.

The Euclidean continuation of the static background exhibits conical singularities which must be regularized.  There are many possible regularizations, the simplest of which is a disconnected geometry with periodic Euclidean time.  This is compatible with quantum mechanics if we consider it as the representation of a thermo-field double (TFD) state.   The temperature is $r_S^{-1}$ in units of the cosmological length scale, and this parameter is fixed by the classical background.  The resulting functional integral is a product of two equal factors. In a sensible quantum theory, this must be interpreted as the square of the partition function of the model\cite{hj,jw}. There are also regularizations with more complicated topology, the simplest of which is a two-sided trumpet, with a cylinder connecting the two Euclidean boundaries.  These geometries have moduli which enable one to move the points of attachment of the  cylinder over the two hyperbolic planes.  Coleman\cite{coleman,coleman2,gs} argued that locality requires also allowing geometries with cylinders connecting arbitrarily many points on the manifold, and that the sum over topologies could be replaced by an averaging of local theories with some probability distribution for couplings, see also~\cite{tblandscape}.  This set of ideas has been recently validated by the remarkable results of\cite{sss}.  Our aim in this paper is to described a particular theory, so we restrict attention to the disconnected topology.  Harlow and Jafferis\cite{hj} have shown that straightforward quantization of pure JT gravity does not satisfy the factorization requirement.  In our model, where the solution space of JT gravity is not quantized, factorization is satisfied and we can view the Euclidean path integral as defining the TFD for a set of density matrices we describe below.

As usual, the Euclidean functional integral for a theory including gravity must be expanded about some fixed configuration of the gravitational fields $S$ and $g_{mn}$, which we take to be a smooth classical solution of JT gravity, with the topology of two copies of the Poincare disk.  
The integral over fluctuations $\Delta S$ is a $\delta$-functional which fixes the metric on the disk to be Euclidean AdS$_2$ globally.  The functional integral over the fermions produces terms in the effective action for $g_{mn}$ that do not depend on $S$:
\begin{align}
S_{eff} = \int d^2x \sqrt{g} \left(a + b R\right)+ \frac{N}{24} \int\int d^2xd^2y\sqrt{g} R (x)\nabla^{-2} (x,y)  \sqrt{g} R(y) .
\end{align}
Comparing with Eq.~(\ref{eq:lag}), the first term in $S_{eff}$ can be absorbed into the AdS$_2$ radius, the second term can be absorbed into a {\it divergent, constant,  real} shift in $S$, and the third term is a bilocal Liouville interaction with fixed coefficient $N/24$ for $N$ fermion fields~\cite{polyakov}.   Written on a generic gravitational background the Liouville term reproduces the conformal anomaly. If we perform the path integral over the $S$ field, then the Liouville term is just a constant, depending only on the moduli of the constant curvature geometry. 

Applying a cutoff to the divergent shift in $S$, we write the vacuum solution as
\begin{equation} S(r) = S_0 - \mu^2 (r - r_s). \label{eq:Swithminussign}\end{equation}
After we consider shock waves below, we will interpret $S$ as the entropy in a causal diamond with right tip  located on a cutoff surface at $r_c$ close to the infinite boundary and with left tip at at $r$. The cutoff surface is related to $S_0$ via $S(r_c)=0$, or $ r_c = S_0/\mu^2 + r_s$.

If, instead, we look at stationary points of the effective action with respect to metric variations for fixed $S$, the Liouville term adds a source to the massive Klein-Gordon equation for $S$.  The source is constant on  the Poincare disk and simply adds a constant to  $S$.  Typically this semiclassical analysis is justified in the large $N$ approximation, but the $\delta$-functional in the path integral over metrics is still enforced by the integration over $S$ fluctuations, so higher-order corrections in the $1/N$ expansion appear to vanish.   This is consistent with the operator solution of the model for arbitrary $N$ described above.  In that solution, the metric is a c-number-valued function, and the operator part of the dilaton field is entirely determined in terms of the $T_{++}$ and $T_{--}$ components of the stress tensor of free fermions in the AdS$_2$ geometry.  The dilaton also has a c-number piece, which is a solution of the massive KG equation with constant source.

It is important to note that the effective action is quadratic in $S$.  Since the metric is not integrated over in our disconnected topological sector, there are no Fadeev-Popov ghosts and no complicated multi-point functions.   As a consequence, the large $N$ expansion truncates, consistent with our exact operator quantization of the model.

\section{Shock Wave Solutions and Entropy}
\label{sec:qnm}
\subsection{Classical Theory}
In order to obtain a better understanding of the physics of our system we now introduce nonzero matter stress-energy in the classical theory. We bosonize the Dirac fermions to scalar fields $\varphi$ and consider coherent states of the bosons.  We begin the analysis in global conformal coordinates and then reinterpret it in static coordinates. In conformal gauge the classical solutions for the scalars are left and right-moving waves $\varphi = f_L(y^-)+f_R(y^+)$.  At $y=\pm \pi/2$, we take boundary conditions $\partial_\tau\varphi=0=(\partial_++\partial_-)\varphi=(f_L'+f_R')$. So the collective coordinate of a $\varphi$ wave packet evolves as $y(t)=y_{0} \pm \tau$ and reflects with inversion off of the boundaries. The metric remains AdS$_2$.  We view the classical evolution of these solutions as the analog of quasi-normal modes of a black hole horizon.

\begin{figure}[t!]
\begin{center}
\includegraphics[width=0.45\linewidth]{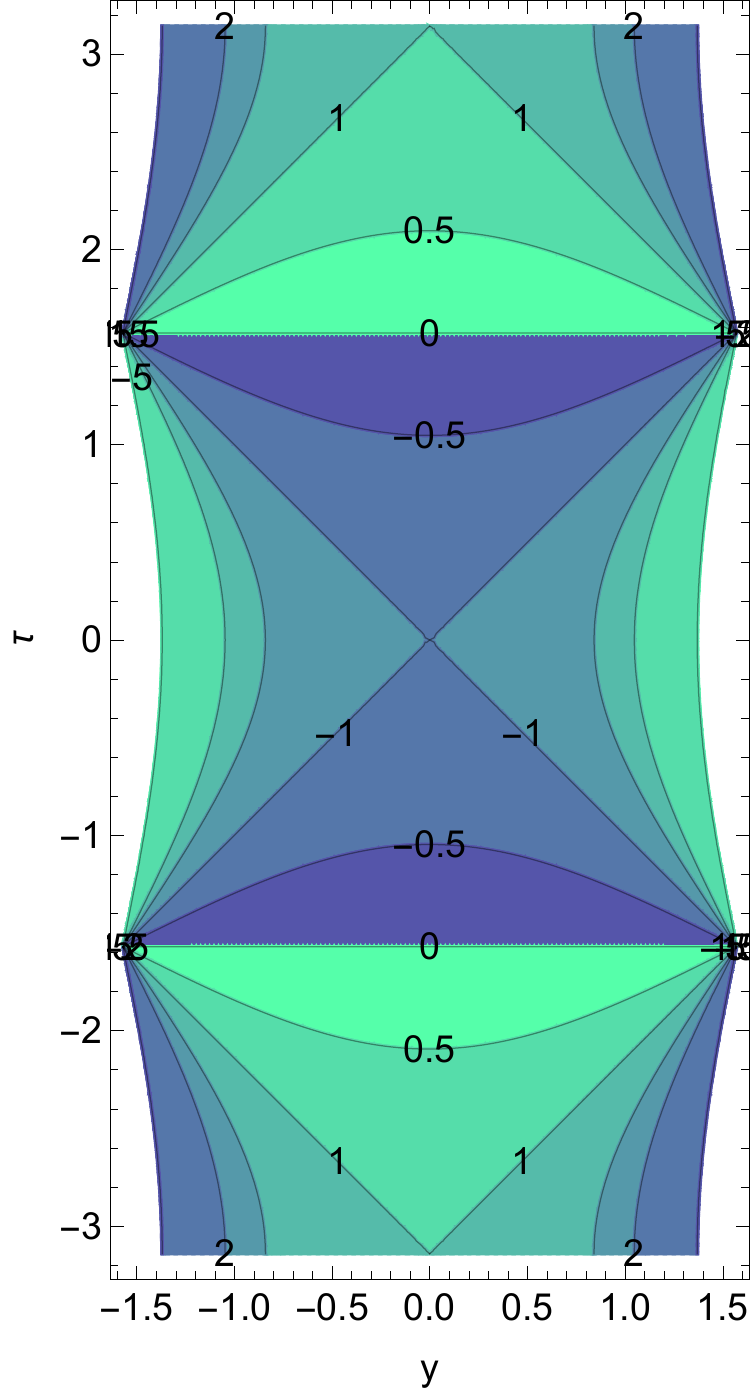}~~~~
\includegraphics[width=0.45\linewidth]{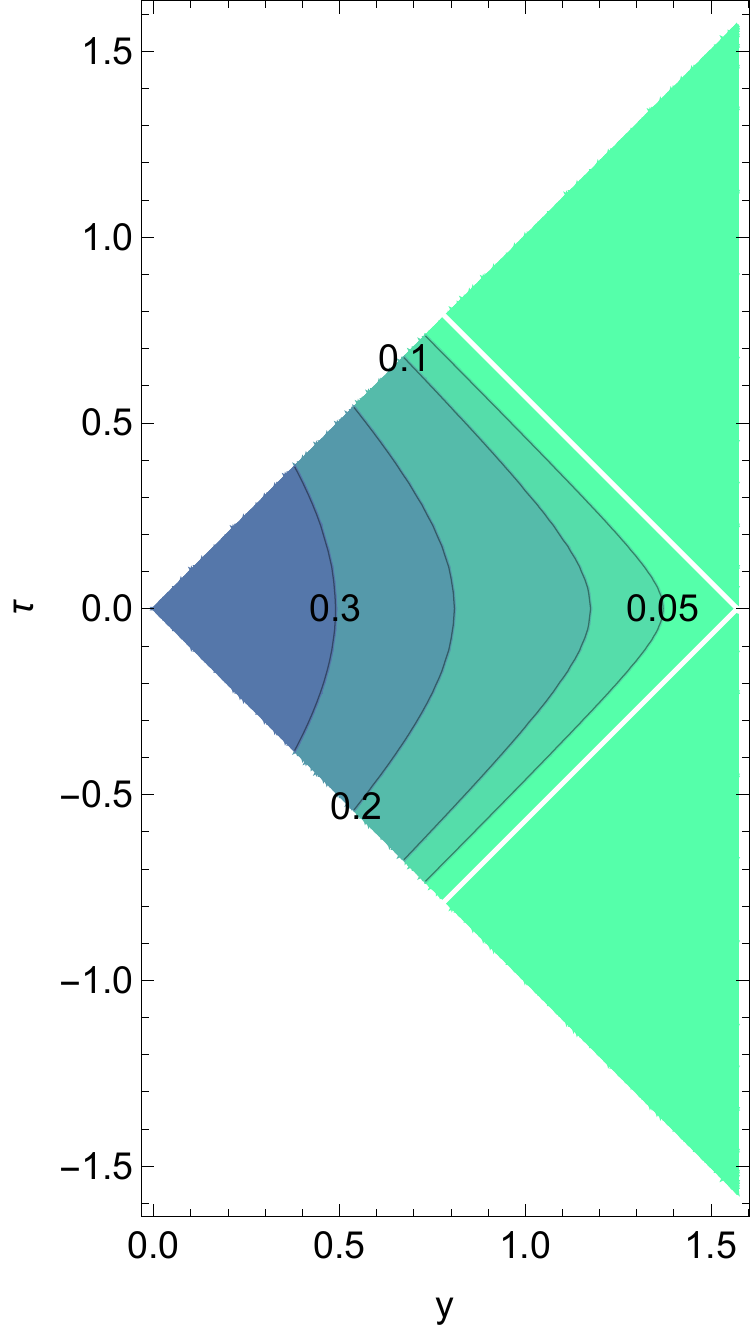}
\caption{Left: Contours of constant $S/\mu^2$ in the vacuum solution, $S/\mu^2=-r$, in global conformal coordinates. Here the infinite constant renormalization is ignored. Right: the change in $S$ induced by a $\varphi$ stress-energy shockwave reflecting off the boundaries (the term proportional to $a$ in Eq.~(\ref{eq:classicalshock}). Here  $a=1$ and we zoom in to show only the right causal diamond. The white line shows the trajectory of the wavefront. After the right-moving wave passes a point $r$, $S$ increases. This is interpreted as energy entering causal diamonds with right tips near the boundary, leading to an increase in their entropy.} 
\label{fig:shockwave}
\end{center}
\end{figure}

What happens to the $S$ field? Following CGHS, let us study a shock wave. The wave that bounces off the boundaries has stress tensors:
\begin{align}
T_{++}&=a\sum_{n=-\infty}^{\infty}\delta(y^+-y_0^+-2n\pi )\nonumber\\
T_{--}&=a\sum_{n=-\infty}^{\infty}\delta(y^--y_0^+-(2n+1)\pi).
\end{align}
At any given $\tau$, only one of the $\delta$ functions is active. 
During a right-moving phase, the solution for the dilaton that satisfies the constraints, behaves like a retarded Green function, and reduces to the dilaton vacuum solution $S=-\mu^2r$ for $a=0$ is:  \begin{align}
S&=-\mu^2\sec{y}\cos{\tau}-\frac{1}{2}a\sec{y}\big[  \theta(y^+-y^+_0-2n\pi )\left(\sin{(y)}-\sin{(\tau-y^+_0)}\right)\big]
\end{align}
During a left-moving phase, the analogous solution is:
\begin{align}
S&=-\mu^2\sec{y}\cos{\tau}+\frac{1}{2}a\sec{y}\big[ \theta(y^--y^+_0-(2n+1)\pi)\left(\sin{(y)}-\sin{(\tau-y^+_0)}\right)\big]
\end{align}
Putting them together, we have
\begin{align}
S&=-\mu^2\sec y\cos\tau -\frac{1}{2}a\sec(y)[\sin(y) - \sin(\tau-y_0^+)]\times\nonumber\\
&~~~~~~~~~~~~~~~~~~~~~~~~~\sum_{n=-\infty}^{\infty}\left( \theta(y^+-y^+_0-2n\pi )-\theta(y^--y^+_0-(2n+1)\pi)\right)
\label{eq:classicalshock}
\end{align}
This solution is plotted in Fig.~\ref{fig:shockwave}. 

\subsection{Quantum Field Theoretic Corrections}
Integrating out  $\varphi$ fluctuations, we obtain one-loop contributions to $\langle T_{\mu\nu}\rangle$ which correct the dilaton EOM and the constraints.
After absorbing an infinite shift into S we can write these as
\begin{align}
\partial_+\partial_- S + \frac{1}{4} e^{2\sigma}S&=-\frac{N}{12}\partial_+\partial_- \sigma\nonumber\\
\partial_+^2 S -2\partial_+\sigma \partial_+ S&=\frac{N}{24}\left(-\partial^2_+ \sigma +(\partial_+\sigma)^2+c_{++}(y_+)\right)+T_{++}\nonumber\\
\partial_-^2 S -2\partial_-\sigma \partial_- S&=\frac{N}{24}\left(-\partial^2_- \sigma +(\partial_-\sigma)^2+c_{--}(y_-)\right)+T_{--}
\label{eq:eomconstr}
\end{align}
$c_{++}$ and $c_{--}$ must be fixed by the boundary conditions on the total stress tensor.
Setting $\sigma$ to its classical value, the  right hand sides of the constraint equations contain the combinations
\begin{align}
-\partial^2_+ \sigma +(\partial_+\sigma)^2 = -\partial^2_- \sigma +(\partial_-\sigma)^2= -\frac{1}{4}. 
\end{align}
while the dilaton equation of motion in conformal static coordinates is
\begin{align}
(-\Box  + 2 r_s^2 {\rm csch}^2(r_s s) )S = \frac{N}{12} r_s^2 {\rm csch}^2(r_s s) 
\end{align}
where $\Box = -4\partial_+\partial_-=-\partial_t^2+\partial_s^2$. 
In static coordinates, measuring $r$ and $t$ in units of $r_s$,
we have
\begin{align}
\left[\partial_t^2  - (1-r^2)^2\partial_r^2 +2  r (1-r^2)\partial_r - 2 (1-r^2)\right]S(t,r) = -\frac{N}{12} (1-r^2) .
\end{align}
For static $S(t,r)=S(r)$, the dilaton EOM is solved by
\begin{align}
S(r) = \frac{N}{24}+c_1 r + c_2 \left(r\tanh^{-1}(r) -1 \right)
\end{align}
where $c_{1,2}$ are constants. The constraint equations are  satisfied for 
\begin{align}
c_2=0, ~~~ c_{--}=c_{++} = 0
\end{align}
again corresponding to the vacuum solution, for which we took $c_1=-\mu^2$. Note that the $N/24$ is another constant contribution to $S$ and can just as well be absorbed into the infinite shift.

For the right-moving shock wave,  $T_{++}^\varphi=a\delta(y^+-y_0^+)$ and $T^\varphi_{--}=0$, and again we  take 
\begin{align}
&c_{--}= 1/4,\nonumber\\
&c_{++}= 1/4.
\end{align}
The constraint equations admit the same solution as previously:
\begin{align}
&S=\frac{N}{24}+\bigg[-\mu^2\sec y\cos\tau -\frac{1}{2}a\sec(y)[\sin(y) - \sin(\tau-y_0^+)]\times\nonumber\\
&~~~~~~~~~~~~~~~~~~~~~~~~~\sum_{n=-\infty}^{\infty}\left( \theta(y^+-y^+_0-2n\pi )-\theta(y^--y^+_0-(2n+1)\pi)\right)\bigg]
\end{align}
So the plots in Fig.~\ref{fig:shockwave} apply also to the quantum corrected case, up to a constant shift of $S$.

At late times we can view these classical solutions as quasi-normal modes of the horizon, and we can understand their most important properties by thinking about the propagation of the classical stress energies $T_{++}$ and $T_{--}$.  These propagate as freely moving waves in the static conformal coordinates.   (These are like tortoise coordinates for a generic black hole; the radial coordinate behaves like
$s \sim r_s^{-1}\log (r - r_s)$ near the horizon and approaches zero at infinity.)  We can consider both right-moving waves that are at an $O(1)$ distance from the horizon at some time $- t_0$ and left-moving  waves that are localized near infinity at that time.  Or we could have a linear combination of the two.   In either case, by a time $T - t_0$ with $T\sim O(1)$, the wave will approach the region where $r - r_s < 1$ (in units of the AdS radius), where dramatic red shifting begins to occur.  Notice that the time at which this occurs is shorter for larger $r_s$ (lower temperature).  

Fig.~\ref{fig:shockwave} shows that the value of $S$ increases (decreases) near the horizon after an outgoing (ingoing) shock wave has passed.  This can be understood from the constraint equation in static conformal coordinates,
$\nabla_+^2 S = T_{++} $, applied to the past boundary of any causal diamond with left tip on the horizon.   The classical stress tensors all satisfy the null energy condition so $\partial_+ S = \int dx^+ T_{++}$ is positive at the horizon.  This will certainly remain true in the quantum field theory as long as the classical contribution dominates the quantum expectation value.  In fact, since the expression involves only the integrated stress tensor, the average null energy condition \cite{anec} which is a rigorous theorem in CFT, is sufficient to prove that the expectation value of $\partial_+ S$ is positive.

\begin{figure}[t!]
\begin{center}
\includegraphics[width=0.45\linewidth]{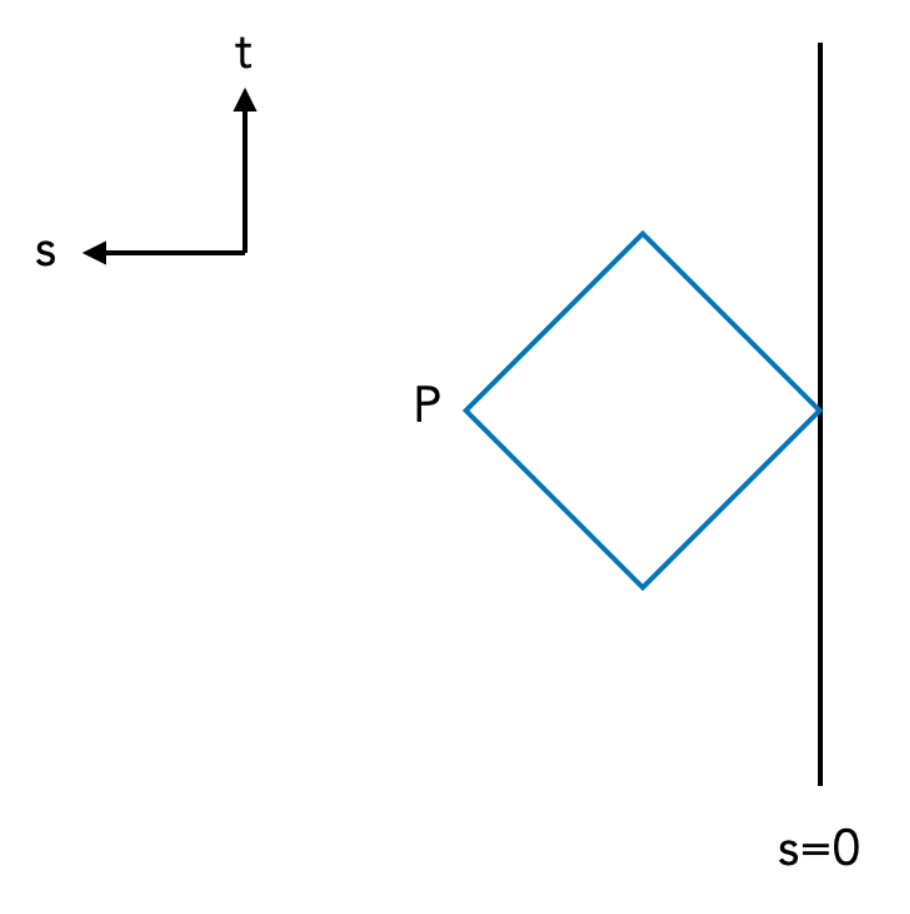}
\caption{In conformal static coordinates the boundary is at $s=0$ and $s$ increases toward the horizon, oriented toward the left in this diagram.  Given a point $P$ in the bulk we draw the diamond with left tip at $P$ and right tip at the same static time on the boundary  (actually, just inside the boundary at a cutoff radius $r=r_c$ as described in the text.) The dilaton value at $P$ (which is independent of $t$) is interpreted as the entropy in this diamond.} 
\label{fig:diamond}
\end{center}
\end{figure}

These remarks support a speculation about both the meaning of the infinite renormalization of the constant piece of $S$, and the $r$ dependence of the $c$-number part of $S$ obtained from the  classical and quantum corrected solutions of the field equations.
A static solution of the field equations represents an equilibrium state of the quantum system.  However, the non-zero value of $T_{+-}$ in the quantum theory proves that the system Hawking radiates.  As in all situations with AdS asymptotics, Hawking radiation bounces off the boundary at infinity, and {\it can} be absorbed by the black hole.  In the AdS$_2$ wormhole geometry, it {\it must} be absorbed by the black hole.
The black hole temperature is fixed, but the entropy of the equilibrium state can change. As mentioned previously throughout this work, we propose an interpretation of $S(r)$ as the entropy in a causal diamond, one of whose spatial tips is at $r$.  A sketch is shown in Fig.~\ref{fig:diamond} in conformal static coordinates.
Thinking of the shockwave solutions as quasi-normal perturbations, we see that energy is transported to the boundary, and reflects off in finite time.  On the other hand it takes an infinite time to propagate to $r = r_s$.   Thus, the equilibrium situation to which any perturbation evolves has all of the entropy collected near the horizon at $r = r_s$.  We therefore rewrite the classical solution for $S$ as in Eq.~(\ref{eq:Swithminussign}),
$ S(r) = S_0 - \mu^2 (r - r_s) $.  The constant $S_0 + \mu^2 r_s$ is, as we have seen, subject to infinite renormalization in quantum field theory. We interpret this as the entropy in a causal diamond whose right tip is on the infinite boundary and whose left tip is at $r$.  If $S_0$ is finite, this equation makes sense only up to a cutoff radius,
\begin{align}
r_c \equiv S_0/\mu^2 + r_s.
\end{align}

In the $AdS_d/CFT$ correspondence for $d \geq 3$, the boundary theory is derived from a picture of branes embedded in asymptotically flat space.  Correlators of boundary operators encode scattering amplitudes of low energy flat space particles from a large stack of branes.  Thermal states  of the CFT  describe near extremal excitations of finite energy density on the branes.  JT gravity is an effective theory of the throat of an extremal black hole in four dimensional gravity, obtained by dimensional reduction of the four dimensional Einstein-Maxwell equations, and linearization of the dilaton.  The authors of\cite{mms} showed that near extremal perturbations deform the $AdS_2$ geometry drastically, so there is no analogous formula for perturbations of the 4D black hole in terms of boundary correlators in $AdS_2$.   The standard treatment of JT gravity shows that the boundary geometry is in fact a fluctuating variable, rather than the frozen geometry that characterizes higher dimensional AdS spaces.  We view the association of finite entropy with a cutoff on the AdS geometry as a further symptom of the difference between $AdS_2$ and higher dimensions.   In our opinion, it means that the dual theory should not be thought of as living on the infinite boundary of the space.   Instead we will interpret the dual theory as located in the vicinity of the horizon $r = r_s$ where the classical perturbations all reside after a finite time.  Correspondingly, our entropy function vanishes at a finite large value of $r$ determined by $S_0$ and reaches its maximum at the horizon $r_s$. 

\section{The Inverted Oscillator and Non-Relativistic Fermions}
\label{sec:IHO}
The key to our UV complete theory of the semi-classical QFT model of fermions coupled to JT gravity is an exact mapping, discovered in\cite{AKK}, between non-relativistic (NR) fermions in the inverted oscillator potential $V = - \frac{1}{2} \omega^2 z^2$ and Weyl fermions scattering off a barrier in Minkowski space. In our model, the geometry is only conformal to Minkowski space, and we apply the equivalence to the conformal static coordinates.  Furthermore, semi-classical considerations indicate that the system is at finite temperature in conformal static coordinates.  We will interpret this in the NR model as a density matrix with a position dependence temperature, and we will find that in the semi-classical approximation to the NR single fermion dynamics, this ansatz can be used to fit the relation between entropy and transit time in the geometrical picture.

The energy scale in the NR model is $\omega^2$ and we will set it equal to $1$.  This is roughly the same scale defined by setting the AdS radius equal to $1$ in JT gravity.  The model can be defined in terms of a pair of canonical variables
\begin{equation} \hat{u}=\frac{\hat{p}_z+\hat{z}}{\sqrt{2}},\ \  \hat{v}=\frac{\hat{p}_z-\hat{z}}{\sqrt{2}},\ \  [\hat{u},\hat{v}] = i ,\end{equation} with single fermion Hamiltonian
\begin{equation} H = \frac{1}{2} (\hat{u}\hat{v} + \hat{v}\hat{u}) . \end{equation}
Note that the operators $\frac{1}{2} \hat{u}^2$,$\frac{1}{2}\hat{v}^2$, $\frac{i}{2}H$ generates an $SL(2,R)$ algebra, which is the algebra of asymptotic symmetries of $AdS_2$.  The Hamiltonian of the model can be written either in terms of fields in the $u$ basis, or in the $v$ basis
\begin{equation} H = \int du \psi^{\dagger} (u) i^{-1} (u\partial_u + \frac{1}{2}) \psi (u) . \end{equation}
\begin{equation} H = \int dv \chi^{\dagger} (v) i (v\partial_v + \frac{1}{2}) \chi (v) . \end{equation}
Going to the variables $x={\rm ln}\, (u,v) $ and defining
\begin{equation} \hat{\psi} (u) = u^{1/2} \psi (u), \ \ \ \ \hat{\chi} (v) = v^{1/2} \chi (v) , \end{equation} we recognize this as the Hamiltonian of left or right moving Weyl fermions.  The two representations are related by a unitary operator. 

The original application of this transformation was to the exact solution of Type 0B string theory.  In that case, the single particle Hilbert space\cite{moore} is defined by scattering boundary conditions at infinity and the unitary transformation above is more or less the scattering operator.  The phrase ``more or less" refers to the fact that the transformation to Minkowski coordinates is singular when $u$ or $v$ vanishes and so there are two kinds of scattering states, related by reflection of the oscillator coordinates $(u,v) \rightarrow - (u,v)$.  The even and odd combinations refer to fermions whose bosonic potentials are the $NS$ and $RR$ tachyons of the $0B$ string.

To interpret the same system in terms of JT gravity, we first gauge the $Z_2$ reflection symmetry, allowing only even single particle states.   We then note that the classical transit time in the oscillator problem is
\begin{equation} T = {\rm ln} \frac{u(T)}{u(0)} . \end{equation} This goes to infinity with $u(T)$ and is finite if $u(T)$ is zero.  Comparison to the transit time in static coordinates for a massless particle in the $AdS_2$ geometry, this suggests that we identify $u = \infty$ with the horizon $r = r_s$ and $u = 0$ with the boundary at infinity.  Our semi-classical investigation suggests that the quantum state of the system represented by a given static classical solution is thermal equilibrium for a detector that uses the static $(t, r)$ coordinate system.   This is not the same as thermal equilibrium in the static conformal $(t,s)$ coordinates.  Rather, a fixed temperature in $AdS_2$ corresponds to a position dependent temperature in the Minkowski coordinates.  Since these coordinates map directly to the oscillator coordinates under the AKK map, this implies a position dependent temperature for fermions in the oscillator potential.

To define what we mean by this, consider a modular Hamiltonian for NR fermions in one spatial dimension $z$, 
\begin{equation} K = \int dz dz^{\prime} \psi^{\dagger} (z) k (z,z^{\prime}) \psi (z^{\prime}) . \end{equation} The log of the partition function and the expectation value of the number operator are given by 
\begin{equation} {\rm ln}\ Z = {\rm Tr}\ {\rm ln}\ (1 + e^{-k}) , \end{equation}
\begin{equation} \langle N \rangle = {\rm Tr}\ \frac{e^{-k}}{(1 + e^{-k})} . \end{equation}

We can take the trace in the over-complete basis of minimal uncertainty states centered at the c-number points $w = (u,v)$ in the classical phase space of the system.
\begin{equation} {\rm Tr}\ O = \int \frac{d^2 w}{\pi} \langle w | O | w \rangle .  \end{equation}
Given the operator $k$ as a Hermitian function of $u$ and $v$ we can write the required traces as integrals over phase space using the Moyal product on phase space functions.  For large values of $u$ and/or $v$ we can approximate the Moyal product by the ordinary product.   We will work in this semi-classical approximation.

We define position dependent temperature by saying that
\begin{equation} k \approx \beta (u) uv , \end{equation} in the semi-classical limit.  The coherent state parameters $u,v$ satisfy the exact quantum equations of motion
\begin{equation} u(t) = u(0) e^t, \ \ \ \ v(t) = v(0) e^{-t} . \end{equation}   These define a transit time for an arbitrary point $u(0), v(0)$ in phase space to the position $U$
\begin{equation} T(U) = {\rm ln}\ (U/u(0)) ,  v(0) = E/u(0) . \end{equation} Our choice of modular Hamiltonian leads us to parametrize phase space in terms of $E, u(0)$ and we have
\begin{equation} du dv = \frac{du dE}{u} . \end{equation}   

Thus, we define a partition function for the system confined to $0 \leq u \leq U$ by
\begin{equation} {\rm ln}\ Z =  \int_0^U \frac{du}{u\pi} \int_0^{\infty} dE{\rm ln}\ ( 1 + e^{-\beta (u) E}) . \end{equation}
We define an average transit time from  a point in phase space, out to some maximal value of $U$, by writing the phase space number density of fermions as
\begin{equation} n(u,E) \equiv \frac{e^{-\beta (u) E}}{1 + e^{-\beta (u) E}} . \end{equation}  The probability of finding a fermion at the phase space point $(u,E)$ is\begin{equation} p(u,E) = \frac{n(u,E)}{\langle N \rangle}. \end{equation}  Note that we are working in the grand canonical ensemble, and assuming $\langle N \rangle\equiv N$ is large so that its fluctuations are small. 
Then the  travel time to $U$, in an average weighted by the probability to be at a phase space point corresponding to the diamond, is\footnote{This is a hypothesis. We could also take the average over the diamond, conditioned on being in the diamond. This changes the factor of $1/N$ in $T(U)$ to $1/N_{in\, diamond}$. This leads to more complicated formulas but not qualitative changes.} 
 \begin{equation} T(U) =  \frac{1}{ N}  \int_0^U \frac{du}{u\pi} \int_0^{\infty} dE\ {\rm ln}\, (U/u)\,n(u,E). \end{equation}  
 We can do the $E$ integrals in both of these expressions, finding
 \begin{equation} S(U) = \int_0^U \frac{du}{u \beta (u)} I_1 .  \end{equation}
 \begin{equation} T(U) = \frac{1}{ N} \int_0^U {\rm ln}\, (U/u) \frac{du}{u \beta (u)} I_2, \end{equation} where 
 \begin{equation} I_1 = \int_0^{\infty} dx\ {\rm ln}\, (1 + e^{-x}) , \end{equation}
 \begin{equation} I_2 = \int_0^{\infty} dx\  \frac{ e^{-x}}{ (1 + e^{-x})} , \end{equation}
 \begin{equation}  N =  \int_0^\infty \frac{du}{u \beta (u)} I_2.\end{equation}
 Note that $I_2$ factors out of the expression for $T$ because we have divided by $ N $.
 It follows that
 \begin{equation} \frac{dS}{dT} = \frac{dS/dU}{dT/dU} = [S\beta(U)]^{-1}  I_1^2 N/I_2 . \end{equation}
 On the other hand we know that in the space-time picture of this system
 \begin{equation} S(r) = S_0 - \mu^2 (r - r_s) . \end{equation}
     \begin{equation} 2r_s T(r) = {\rm ln}\, \frac{r+r_s}{r - r_s} . \end{equation}
 Thus 
 \begin{equation} \frac{dS}{dT} = \mu^2 (r^2 - r_s^2) = (S_0 - S) ([S_0 - S]\mu^{-2} + 2r_s) . \end{equation}
 Matching the two expressions for $dS/dT$ we obtain an equation for $\beta (U)$ . \begin{equation} \beta^{-1} (U) = \frac{I_2}{I_1^2 N} S (S_0 - S) (\mu^{-2} [S_0 - S] + 2r_S).  \end{equation}   Reexpressing this in terms of $dS/dU$, we get a differential equation for $S$ as a function of $U$
 \begin{equation}U\frac{dS}{dU} =   \frac{I_2}{I_1 N}S (S_0 - S) (\mu^{-2} [S_0 - S] + 2r_S).  \end{equation} This can be solved by quadratures, expressing $U$ as a function of $S$, 
 \begin{equation} U = A^{\frac{1}{\alpha (1 + \gamma)}}(1 - A)^{\frac{1}{\alpha  \gamma}}(1 - A + \gamma)^{\frac{1}{\alpha \gamma (1 + \gamma)}},\end{equation}
 where
 \begin{equation} A = \frac{S}{S_0} \ \ \ \ \gamma = \frac{2\mu^2 r_s}{S_0} \ \ \ \ \alpha = \frac{S_0^2 I_2 }{\mu^2 I_1  N} .
 \end{equation} 
 Note that $U (0) = 0$ and $U(S_0) = \infty$.  This tells us that the maximal entropy is reached near the horizon, to which the travel time from the boundary is infinite.  Plugging in the formula $S = S_0 - \mu^2 (r - r_s)$ we get a monotonic functional relation between $U$ and the space-time coordinate $r$.
 
 Our solution of the model is based on the assumption that the maximal entropy $S_0$ is obtained when all of the fermions cluster near the horizon.  This is compatible only with classical solutions of JT gravity with positive $\mu^2$.  We will comment on the meaning of solutions with negative $\mu^2$ when we discuss the relation of this model to four dimensional extremal RN black holes. 
 \subsection{Cutoffs and Additional Interactions}
 
 Via the non-relativistic inverted harmonic oscillator map, we have constructed unitary and complete models for each of the static equilibrium solutions in JT gravity, with positive $\mu^2$. The oscillator phase space coordinate $u$ maps to the spatial static conformal coordinate in the gravity solution. There are, however, two notable defects of the models we've constructed so far. We will now discuss some cutoffs and interactions that can be added to improve the models.
 
   The first defect is not terribly serious: the Hilbert space has infinite maximal entropy.   However, the thermal states we've constructed have a built in ultraviolet cutoff, so insisting on a hard cutoff on the UV spectrum of the inverted oscillator does not affect the results.  We do this in two steps.  First introduce a finite cutoff at a large value of the oscillator spatial coordinate $z = z_c$\footnote{We note that Moore's\cite{moore} careful definition of the scattering states of the oscillator starts from a finite $z$ interval with appropriate boundary conditions.  For our purposes we need reflecting boundary conditions and we keep the cutoff finite. We will find a more elegant version of this cutoff procedure below, by returning to its description as a limit of a finite dimensional matrix quantum mechanics.}.  This discretizes the spectrum.  We then make the Hilbert space finite by imposing a cutoff on the magnitude of the eigenvalue of the Hamiltonian.   
 
 A more serious problem is the fact that the model is completely integrable, and so does not contain a mechanism for thermalizing the system, although semi-classical analysis indicates that the model has only modified thermal equilibrium states.   This defect can be repaired by adding random couplings to the models at $z_c$.  Since all single particle states enter that regime after a finite time, those couplings will thermalize any initial state.  If they are chosen to be typical members of the SYK ensemble, then the low temperature partition function will be of the form calculated from the Euclidean path integral of JT gravity\cite{SYK}.  
 
 These cutoffs spoil the elegant mapping to relativistic fermions, but for large $z_c$ this only occurs in a small region near the horizon.   We've already remarked that the thermal nature of our density matrix implies that our results are insensitive to an eigenvalue cutoff.
 With these modifications we believe that the models fully describe the quasi-normal mode solutions which exhibit the local space-time structure, because relativistic fermion dynamics away from the horizon maps to the regime of phase space in the NR model far from the cutoffs where the random interactions reside.  The near horizon behavior of QNMs is, as usual, evocative of return to thermal equilibrium, and  the interactions above assure that equilibration occurs.
 
 The quantum states we have studied have the same entropy vs. transit time relation as causal diamonds in the space-time picture, with $S(r)$ interpreted as the entropy.\footnote{We remind the reader that there was an ambiguity in defining the probability of finding a fermion in a particular causal diamond, depending on our choice of the denominator.  This modifies the functional form of $\beta(U)$.}  
 These quantum states were constructed in the semi-classical approximation to the single particle non-relativistic quantum mechanics problem.  The metric and dilaton of JT gravity are constructed as hydrodynamic variables which describe coarse grained properties of the underlying fermion system. They are not independent dynamical variables.   It's interesting to note that the parameters of the JT gravity solutions are encoded in the state of the fermion system through the function $\beta (U)$.  One of the authors has emphasized\cite{tblandscape} that in the context of asymptotically flat or AdS models of string theory, backgrounds with different asymptotic behavior correspond to different quantum Hamiltonians, rather than different states in a single quantum model.  Those arguments were all related to the high energy behavior of the system and the properties of very large black holes.  The JT gravity models have only a finite entropy, and our study of shock waves showed that perturbations of the equilibrium states $S_0, \mu^2, r_s$ could change the value of $S_0$.  We believe that this is an artifact of the effective field theory description, in which $S$ gets infinite renormalizations.  Once we impose cutoffs, $S_0$ is the maximal entropy of the system, and cannot be varied without changing the model.

\subsection{Four dimensional near-extremal black holes}
\label{sec:RNextremal}

JT gravity and the associated SYK models were originally proposed as near-horizon limits of extremal RN black holes in 4D. The extremal black hole has metric 
\begin{align}
 ds^2 = -(1-r_s/r)^2 dt^2 + (1-r_s/r)^{-2} dr^2 + r^2 d\Omega_2^2.
 \label{eq:extremalRN}
\end{align}
This becomes $AdS_2 \times S^2$ in the leading near-horizon expansion, i.e., up to corrections at order $r/r_s-1$. 
JT gravity is obtained from by dimensionally reducing Einstein-Maxwell theory in a certain frame and then expanding the dilaton to linear order around a constant. The  $\mu=0$ solution of the JT theory corresponds to the $AdS_2 \times S^2$ near-horizon limit of extremal RN.   We note that if we consider the next order terms in the expansion of the metric around its extremal throat, $S$ behaves like the linear solution of JT gravity {\it with negative $\mu^2$}, corresponding to the growth of two spheres as we recede from the throat.

It was pointed out in~\cite{mms} that, in stark contrast to higher-dimensional AdS/CFT correspondences, near-extremal black holes do not have a description in terms of excitations of the extremal system.  Arbitrarily small perturbations of the extremal solution replace the infinite throat by a space-like singularity.  Instead, the AdS$_2$ effective theory describes only the dynamics of the classically-degenerate ``zero" temperature limit of the model.
The classical geometry does not tell us how (or if) the apparent degeneracy is lifted. 

JT gravity path integrals, which agree with the spectral density of the SYK models, imply that the classical degeneracy is lifted and the model has a complicated spectrum of low energy levels.  Our model reflects similar properties.  It has the advantage that the model  can describe short-lived local excitations in AdS$_2$, before they fall to the horizon.  For example, our quasi-normal mode solutions describe states in the Weyl fermion Hilbert space, and so can be translated into states in the IHO Hilbert space.  If they do not involve large energies, then they will be states in the cutoff model, which evolve like states in quantum field theory until they get close to the horizon. So far as we know, there is no similar interpretation of the SYK model in terms of local physics.\footnote{The papers of~\cite{grossrosenhaus} display the boundary correlators of the SYK models in terms of Witten diagrams of an infinite number of scalar fields in Euclidean $AdS_2$, but there is no gap in the spectrum of fields that would allow for an interpretation in terms of local physics.}  Susskind\cite{falling} has attempted to map certain kinematic properties of single particle excitations moving in $AdS_2$ into the time evolution of single fermion operators in the SYK model, but this is much less than one would expect from effective field theory in the low curvature throat of a large charge extremal black hole.

The absence of near-extremal excitations might be associated with the fact that 
none of these models contain degrees of freedom associated with the 2-sphere in the true AdS$_2 \times S^2$ near-horizon region. Near extremal black holes can carry angular momentum, and also contain states in which localized excitations penetrate the 2-sphere in angularly localized regions.  There is nothing in our model that corresponds to such excitations, or to the phenomenon exhibited in\cite{mms}, in which a local perturbation of the extremal black hole destroys the $AdS_2$ geometry.  

Thus, we believe the solutions of JT gravity with negative $\mu^2$ do not represent physics in the extremal black hole geometry.  We cannot probe the small region of the extremal geometry where they are a good approximation, without drastically modifying the geometry. Solutions with positive $\mu^2$ have the entropic interpretation described above, but should not be thought of as reflecting the leading classical deviation from the near-horizon limit.  The solution with $\mu^2 = 0$  represents the extremal RN geometry in models of 4 dimensional asymptotically flat space where these solutions are BPS states.  In that case supersymmetry tells us that the zero temperature entropy arises from an exact vacuum degeneracy and the global $AdS_2$ geometry, with no Einstein-Rosen bridge, describes the dimension of the degenerate Hilbert space.  Again, excitations of the BPS black hole, are not included in this Hilbert space.

To end this section we want to summarize the logic of our quantum model and the way it maps to the field theoretic model of fermions coupled to JT gravity.
\begin{itemize}
\item We write the entropy on the gravity side of the duality as $S_0 - \mu^2 (r - r_S)$.  We interpret this as the entropy in a causal diamond with one spacelike tip at $r$ and the other at a point near infinity where this function vanishes.  The basis for this interpretation on the gravity side is our calculation of what happens to coherent states of the bosonized fermions, which we take to be indicative of what will happen to arbitrary fermion states:  for most of static time the solutions hover very close to the horizon, but due to Hawking radiation there is always some probability for fermions to be anywhere in the bulk.   On the NR side, the horizon is mapped (in the WKB/Thomas-Fermi approximation to the NR fermion problem) to oscillator coordinate $z = \infty$.  We construct (in the same approximation) density matrices whose entropy can be localized in the $u = z + p_z $ coordinate and have maximal entropy $S_0$ when the fermions are at $z = \infty$.  This gives a map from the $u$ coordinate to the $r$ coordinate on the gravity side.   The map is tied down by insisting that the travel time from the boundary at $r = r_s + S_0/\mu^2$,  calculated from geometry on the gravity side, is the same as the average travel time calculated from the density matrix from a generic starting value of u to infinity on the oscillator side.   The spacetime coordinate $r$ is thus emergent on the NR fermion side: to define it we need to use the WKB approximation.  $ t$ on the other hand is simply identified on both sides because of time translation symmetry of the problem.  
\item The AKK map is a local map between r on the gravity side and (either $v$ or) $u, (r = \ln\,u) $ on the oscillator side.  The geometry does not define a scattering theory, in the JT case.  We deal with this by the following set of rules:
\item We gauge the $y$ reflection symmetry, allowing only even wave functions.  This means we deal with only one sign of $y$ or $u$.
\item We use minimal uncertainty states to define quantum phase space coordinates that satisfy exact equations of motion $E = const$, $u = u(0)e^t$.   To do things exactly on the NR side we would have to use Moyal products of phase space functions.  Our WKB approximation consists of replacing these by ordinary products.  The commutator between $E$ and $u$ is exactly given by $i$ times the Poisson bracket, and apart from that we treat phase space functions as commuting.  This is the approximation in which a classical space time description is emergent.
\item Finally, in order to make the Hilbert space finite dimensional (as expected for a 4d extremal black hole), we cut off both the oscillator coordinate and the absolute value of the single particle energy.   Note that Moore\cite{moore} derived the scattering description for the fermions by starting with a finite $z$ coordinate and taking the limit with certain boundary conditions. We just use reflecting boundary conditions at $z_c$, instead of the ones that lead to scattering theory. The UV cutoff does not affect our previous considerations because all of our density matrices cut off high energy states anyway.  SYK interactions localized at $z_c$, plus the fact that our density matrices have most of the probability concentrated there imply that the partition function of the model will look like that of the SYK model.   The advantage of our model is that we have an explicit description of evanescent states localized far from the horizon, which we expect to have for an extremal 4D black hole of large charge.
\end{itemize}

\section{Matrix Model Interpretation}

The non-relativistic fermion model that we have employed is not obviously holographic.  One has to remember its origin as a limit of a large $N$ matrix quantum mechanics~\cite{Brezin:1977sv} in order to see the underlying holography.\footnote{The authors would like to thank L. Susskind for suggesting the addition of this section to our paper.   Susskind\cite{Susskind:2021omt} has previously used the cutoff inverted oscillator model, with chaotic interactions added at the coordinate cutoff, as a toy model for de Sitter space.}  We will briefly recall this and see that the cutoffs we have imposed are a natural result of keeping $N$ finite.  Consider matrix quantum mechanics with a single-trace action:
\begin{align} S =N \int dt\, {\rm Tr }\, [\dot{M}^2 / 2 - V(M)] , \end{align} 
with $V(M)$ a double well potential.  Write $M = U^{\dagger} \Lambda U $, where $U$ is unitary and $\Lambda$ is diagonal.  This representation has a gauge redundancy, which permutes the eigenvalues of $\Lambda$.  The integral over $U$ can be done exactly and produces the Van der Monde determinant
\begin{align}
    \prod_{i \neq j} (\lambda_i - \lambda_j)^2 .
\end{align}   
The resulting action is that of non-interacting ``particles" propagating in a one dimensional space, with a single particle potential $V$.  If we ``canonically normalize" by absorbing the square root of the Van der Monde determinant into the initial and final particle states, then the antisymmetry of the Pfaffian  and the permutation gauge invariance imply that the particles obey Fermi statistics~\cite{Brezin:1977sv}.

The double scaling limit\cite{brezinetal}\cite{davidandigor} takes $N$ to infinity (which also takes the number of fermions to infinity), while scaling the Fermi level to sit right at the top of the potential as $N$ goes to infinity.  The AKK map then shows that in this limit we have an emergent space-time, in which massless relativistic fermions propagate.  Since massless fermions are conformally invariant, it is possible to build models of space-times conformal to Minkowski space as well.   The finite $N$ model automatically imposes the cutoff on the eigenvalue coordinate that we imposed by hand in the previous section.  We can place the chaotic interactions at the minimum of the potential.\footnote{We continue to impose the $Z_2$ gauge restriction to even parity wave functions, so there is only one minimum.}  If we add, in addition, an attractive density-density interaction, then equilibrium states at the minimum will have large negative energies for large $N$ and most of the high energy spectrum of the model will be unreachable by finite perturbations of the equilibria.  We are free to cut it off in any way we like.    From the point of view of the matrix model, four fermion interactions are double trace operators.    Thus, our model is a holographic $0+1$ dimensional quantum mechanics of matrices, dual in the large $N$ double scaling limit to a model of relativistic fermions interacting with gravity in $1 +1$ dimensions.

   \section{Conclusions}
   
   We have shown that two dimensional dilaton gravity theories can be derived as covariant versions of Jacobson's\cite{ted} hydrodynamic equations for an entropy function for the holographic screen of causal diamonds (which is two disconnected points in two dimensions).   These equations contain a covariantly-conserved stress tensor, which is supposed to be the expectation value of some quantum operator in a near equilibrium state.  We then explored a quantum theory consistent with these equations, which provides a model for the stress tensor in the hydrodynamical equations.

 We  constructed an exactly soluble quantum field theory model, which reduces in a semiclassical approximation to JT gravity coupled to  free massless fermions.  
 Various coordinate systems are useful for different purposes, and we considered two different  systems, the global conformal  and static conformal coordinates.\footnote{Each of these is related to non-conformal system by a transformation of space coordinates.  It is easy to go back and forth between two spatially related coordinate systems, even in the quantum theory.  The non-conformal systems are those naturally associated with actual observations by geodesic observers.}  Following~\cite{hj,jw} we claimed that the only system in which it made sense to search for equilibrium states was the static one, where there is a time-independent Hamiltonian.  They cut the classical $AdS_2$ manifold down to a wormhole space-time with an Einstein-Rosen bridge connecting two asymptotically $AdS_2$ boundaries.
 
 In our model Lagrangian, fermions interact only with the metric, and not with the entropy field.  We discussed two methods of quantization, Schwinger's old fashioned method of consistency between classical and Heisenberg equations of motion for the operators of the theory, and Euclidean functional integration.  Both led to the conclusion that the $AdS_2$ metric is locally a classical variable, which up to boundary conditions, linearizes the equations of motion and allows for an exact solution.  We showed that the sourced massive Klein-Gordon equation for the entropy field $S$ followed from conventional quantization of the Dirac field, up to a constant shift.   Functional integral quantization leads one to expect an infinite additive renormalization of $S$ and the constant value of $S$ can be absorbed into the two parameters characterizing the classical background solution of pure JT gravity.  Note that all of these manipulations could be done in a completely covariant manner, and so were valid independent of our choice of coordinates and boundary conditions.  
 
 These field theory models contain infinities, which one does not expect in a UV finite theory of quantum gravity.  More seriously, although the only stationary states of the model are thermal, the model is completely integrable and does not exhibit the spectral properties expected in quantum gravity.  There are an infinite number of global conservation laws and a continuous spectrum of fermion states, rather than a discrete chaotic spectrum.

 Using the AKK map between non-relativistic fermions in an IHO potential and relativistic Weyl fermions, we have constructed a UV complete model which exhibits the semi-classical behavior of the field theory model, but can be easily modified to remove its defects.  There are several striking features of our model.  First of all, despite the fact that the geometry of space-time is locally $AdS_2$, it is not correct to think of the dual fermions as operators sitting at the boundary of space-time, forming some deformed conformal quantum mechanics.   Indeed, it is well known that in attempting to interpret the model in this way, one comes to the conclusion that the AdS radius is a microscopic scale.   This is at odds with the derivation of the geometry from the near horizon behavior of extremal charged four dimensional black holes.  Instead we have concluded that ``the hologram sits on the horizon".  Semi-classical calculations indeed show that at asymptotic static times, all excitations approach the horizon.  The semi-classical model also predicts Hawking radiation, which bounces off infinity and returns to the vicinity of the horizon in finite time.  Nonetheless we expect that the static equilibrium solutions of the model correspond to a quantum state in which some number of fermions will exist in every causal diamond anchored on the boundary and some finite interior point.  We parametrize the static solution of the dilaton so that it can be interpreted as the total entropy in such a diamond.  Insisting that the maximal entropy $S_0$ of the system be finite, one finds that one is forced to truncate $AdS_2$ at a maximum radius $S_0 +\mu^2 r_s = r_c . $  
 
 In our semi-classical approximation to the non-relativistic fermion problem the horizon is associated with the line $u = \infty$ in the coherent state basis of the single particle problem.  However, since our density operators have finite entropy for all finite values of the parameters in the model, nothing can be affected by truncating the single particle Hilbert space to make it finite-dimensional. The easiest way to do this is to put reflecting boundary conditions at a finite value of the oscillator coordinate $z$ and restricting the maximal allowed eigenvalue of the single particle Hamiltonian.  We then add interactions between the fermionic oscillators that are retained by the cutoff, which are chosen randomly from the SYK ensemble and localized near the cutoff value of $z$.  This makes the near-horizon dynamics chaotic and its spectral density coincide
 with that calculated from the Euclidean path integral of JT gravity.
 Thus we obtain a model which exhibits both the semi-classical behavior associated with QFT in the throat of an extremal four dimensional black hole of large charge, but also the chaotic quantum dynamics expected from general principles of quantum gravity.   The realization of the IHO fermion model as a double scaling limit of a large $N$ matrix model, exhibits its holographic nature, and naturally incorporates the cutoffs described above.
 
 For fixed finite entropy there are a huge number of different models which have all of these properties.  This should not surprise us.  There are undoubtedly a large number of string models in asymptotically flat space, which have non-BPS extremal charged black holes as excitations\footnote{The model will be supersymmetric, but not all of its extremal black hole excitations are.}.  One can have
 many $U(1)'s$ or many different spectra of charged particle masses under a single $U(1)$.  It should not be surprising that the detailed near horizon dynamics is different in different models.  Furthermore, there is nothing wrong with having models of quantum gravity in two dimensions, which are not approximations to higher dimensional models.  Averaging over the couplings in these models is a good way to extract coarse grained properties of the spectrum.  The folklore of the quantum chaos literature suggests that it is a computationally convenient way to approximate time averaged quantities.   It is not a fundamental property of quantum gravity.

 \section{Acknowledgements}
 PD acknowledges support from the US Department of Energy under Grant number DE-SC0015655. The work of TB and BZ was partially supported by the US Department of Energy under Grant No.  DE-SC0010008.  We thank Juan Maldacena for comments on an earlier version of the manuscript.  We particularly thank L. Susskind for incisive comments that led to the Matrix model interpretation of our results.  The connection with his use of the same quantum model as a toy model of de Sitter space is fascinating and remains to be explored.


\begin{thebibliography}{99}
\bibitem{AKK}S.~Alexandrov, V.~ Kazakov, and I.~ Kostov. "Time-dependent backgrounds of 2D string theory." Nuclear Physics B 640.1-2 (2002): 119-144. [	arXiv:hep-th/0205079]
\bibitem{JT} R. Jackiw, {\it Lower dimensional gravity}, Nuclear Physics B 252 (1985) 343-356; C. Teitelboim, {\it Gravitation and Hamiltonian Structure in Two Space-Time Dimensions},
Phys. Lett. 126B (1983) 41-45; T. Muta and S. D. Odintsov, {\it Two-dimensional higher derivative quantum gravity with
constant curvature constraint}, Prog. Theor. Phys. 90 (1993) 247-255; J. P. S. Lemos and P. M. Sa, {\it Nonsingular constant curvature two-dimensional black hole},
Mod. Phys. Lett. A9 (1994) 771-774, [arXiv:gr-qc/9309023]; J. P. S. Lemos, {\it Thermodynamics of the two-dimensional black hole in the Teitelboim-Jackiw
theory}, Phys. Rev. D54 (1996) 6206-6212, [arXiv:gr-qc/9608016].
\bibitem{SYK} S. Sachdev and J. Ye, {\it Gapless spin fluid ground state in a random, quantum Heisenberg magnet}, Phys. Rev. Lett. 70 (1993) 3339, [arXiv:cond-mat/9212030]; 
S.~Sachdev,
``Holographic metals and the fractionalized Fermi liquid,''
Phys. Rev. Lett. \textbf{105}, 151602 (2010)
[arXiv:1006.3794 [hep-th]];
A. Kitaev, {\it A simple model of quantum holography}, Talks at KITP (2015) ;
J.~Maldacena and D.~Stanford,
``Remarks on the Sachdev-Ye-Kitaev model,''
Phys. Rev. D \textbf{94}, no.10, 106002 (2016)
[arXiv:1604.07818 [hep-th]].
\bibitem{cghsrst} C.~G.~Callan, Jr., S.~B.~Giddings, J.~A.~Harvey and A.~Strominger, {\it Evanescent black holes,}
Phys. Rev. D \textbf{45}, no.4, 1005 (1992)
[arXiv:hep-th/9111056]; J.~G.~Russo, L.~Susskind and L.~Thorlacius,
{\it The Endpoint of Hawking radiation,}
Phys. Rev. D \textbf{46}, 3444-3449 (1992)
[arXiv:hep-th/9206070]; J.~G.~Russo, L.~Susskind and L.~Thorlacius,
{\it Black hole evaporation in (1+1)-dimensions,}
Phys. Lett. B \textbf{292}, 13-18 (1992)
doi:10.1016/0370-2693(92)90601-Y
[arXiv:hep-th/9201074];
\bibitem{lindil1} T.~Banks, {\it Holographic Space-time Models in $1 + 1$ Dimensions,}
[arXiv:hep-th/1506.05777]. 
\bibitem{lindil2} T.~Banks,
{\it Microscopic Models of Linear Dilaton Gravity and Their Semi-classical Approximations,}
[arXiv:hep-th/2005.09479].
\bibitem{trivedi} U.~Moitra, S.~K.~Sake, S.~P.~Trivedi and V.~Vishal,
``Jackiw-Teitelboim Model Coupled to Conformal Matter in the Semi-Classical Limit,''
JHEP \textbf{04}, 199 (2020)
doi:10.1007/JHEP04(2020)199
[arXiv:hep-th/1908.08523];  
U.~Moitra, S.~K.~Sake and S.~P.~Trivedi,
``Aspects of Jackiw-Teitelboim Gravity in Anti-de Sitter and de Sitter spacetime,''
[arXiv:2202.03130 [hep-th]].
\bibitem{ted} T.~Jacobson,``Thermodynamics of space-time: The Einstein equation of state,''
Phys. Rev. Lett. \textbf{75}, 1260-1263 (1995)
doi:10.1103/PhysRevLett.75.1260
[arXiv:gr-qc/9504004].
\bibitem{hj} D.~Harlow and D.~Jafferis, ``The Factorization Problem in Jackiw-Teitelboim Gravity,''
JHEP \textbf{02}, 177 (2020)
doi:10.1007/JHEP02(2020)177
[arXiv:hep-th/1804.01081].
\bibitem{jw}
D.~L.~Jafferis and D.~K.~Kolchmeyer, ``Entanglement Entropy in Jackiw-Teitelboim Gravity,''
[arXiv:hep-th/1911.10663].
\bibitem{kitaevsuh} Kitaev and S. J. Suh, Statistical mechanics of a two-dimensional black hole, JHEP 05 (2019) 198, [arXiv:hep-th/1808.07032].
\bibitem{moore}G.~Moore, {\it Double-scaled field theory at c= 1.} The Large N Expansion In Quantum Field Theory And Statistical Physics: From Spin Systems to 2-Dimensional Gravity 1993(pp. 1054-1087);
P.~H.~Ginsparg and G.~W.~Moore,
{\it Lectures on 2-D gravity and 2-D string theory.}[arXiv:hep-th/9304011].
\bibitem{coleman} S.~R.~Coleman, {\it Black Holes as Red Herrings: Topological Fluctuations and the Loss of Quantum Coherence,}
Nucl. Phys. B \textbf{307}, 867-882 (1988)
doi:10.1016/0550-3213(88)90110-1
\bibitem{coleman2}S.~R.~Coleman,
{\it Why There Is Nothing Rather Than Something: A Theory of the Cosmological Constant,'}
Nucl. Phys. B \textbf{310}, 643-668 (1988)
doi:10.1016/0550-3213(88)90097-1
\bibitem{gs}S.~B.~Giddings and A.~Strominger,
{\it Baby Universes, Third Quantization and the Cosmological Constant,'}
Nucl. Phys. B \textbf{321}, 481-508 (1989)
doi:10.1016/0550-3213(89)90353-2
\bibitem{tblandscape}T.~Banks,
{\it Prolegomena to a Theory of Bifurcating Universes: A Nonlocal Solution to the Cosmological Constant Problem Or Little Lambda Goes Back to the Future,'}
Nucl. Phys. B \textbf{309}, 493-512 (1988)
doi:10.1016/0550-3213(88)90455-5
\bibitem{sss} P.~Saad, S.~H.~Shenker and D.~Stanford,
{\it JT gravity as a matrix integral,}
[arXiv:hep-th/1903.11115].
\bibitem{polyakov}
A.~M.~Polyakov,
``Quantum Geometry of Bosonic Strings,''
Phys. Lett. B \textbf{103}, 207-210 (1981).
\bibitem{sw}D.~Stanford and E.~Witten,
{\it JT gravity and the ensembles of random matrix theory,'}
Adv. Theor. Math. Phys. \textbf{24}, no.6, 1475-1680 (2020)
doi:10.4310/ATMP.2020.v24.n6.a4
[arXiv:hep-th/1907.03363].
\bibitem{anec} T. Faulkner, R. G. Leigh, O. Parrikar and H. Wang, {\it Modular Hamiltonians for Deformed Half-Spaces and the Averaged Null Energy Condition,} JHEP 09, 038, 2016, [arXiv:hep-th/1605.08072];
T. Hartman, S. Kundu and A. Tajdini, {\it Averaged Null Energy Condition from Causality}, JHEP 07, 066, 2017, [arXiv:hep-th/1610.05308].

\bibitem{mms} J.~M.~Maldacena, J.~Michelson and A.~Strominger,
{\it Anti-de Sitter fragmentation,}
JHEP \textbf{02}, 011 (1999)
doi:10.1088/1126-6708/1999/02/011
[arXiv:hep-th/9812073].
\bibitem{grossrosenhaus} D.~J.~Gross and V.~Rosenhaus,
{\it The Bulk Dual of SYK: Cubic Couplings,}
JHEP \textbf{05}, 092 (2017)
doi:10.1007/JHEP05(2017)092
[arXiv:hep-th/1702.08016].
\bibitem{Schwinger}J.~Schwinger, "The theory of quantized fields. I." Physical Review 82.6 (1951): 914; J.~Schwinger, "The theory of quantized fields. II." Physical Review 91.3 (1953): 713.
\bibitem{Susskind:2021omt}
L.~Susskind,
``De Sitter Holography: Fluctuations, Anomalous Symmetry, and Wormholes,''
Universe \textbf{7}, no.12, 464 (2021)
doi:10.3390/universe7120464
[arXiv:2106.03964 [hep-th]].
\bibitem{Brezin:1977sv}
E.~Brezin, C.~Itzykson, G.~Parisi and J.~B.~Zuber,
``Planar Diagrams,''
Commun. Math. Phys. \textbf{59}, 35 (1978)
doi:10.1007/BF01614153
\bibitem{brezinetal}
E.~Brezin, V.~A.~Kazakov and A.~B.~Zamolodchikov,
``Scaling Violation in a Field Theory of Closed Strings in One Physical Dimension,''
Nucl. Phys. B \textbf{338}, 673-688 (1990)
doi:10.1016/0550-3213(90)90647-V
\bibitem{davidandigor}
D.~J.~Gross and I.~R.~Klebanov,
``Fermionic string field theory of c = 1 two-dimensional quantum gravity,''
Nucl. Phys. B \textbf{352}, 671-688 (1991)
doi:10.1016/0550-3213(91)90103-5

\bibitem{falling}
L.~Susskind,
``Why do Things Fall?,''
[arXiv:1802.01198 [hep-th]].

\end{thebibliography}
\end{document}